\documentstyle[12pt,psfrag]{article}
\newcommand{\be}{\begin{equation}}
\newcommand{\ee}{\end{equation}}
\newcommand{\ba}{\begin{eqnarray}}
\newcommand{\ea}{\end{eqnarray}}
\def\L5{\tilde{\Lambda}}

\begin{document}
\newsavebox{\dotdot}
\savebox{\dotdot}[3mm]{\shortstack{\circle*{0.8}\\ \\ \circle*{0.8}}}
\newcommand{\nord}{\usebox{\dotdot}}

\vskip 1cm
\begin{center}
{\Large {\bf Nature of the deconfining 
phase transition in the 2+1-dimensional SU(N) 
Georgi-Glashow model
}}\\[1cm]
P. Lecheminant 
\footnote{Philippe.Lecheminant@ptm.u-cergy.fr},
\\
{\it Laboratoire de Physique Th\'eorique et
Mod\'elisation, CNRS UMR 8089, \\
Universit\'e de Cergy-Pontoise, \\
Site de Saint-Martin,
2 avenue Adolphe Chauvin,\\
95302 Cergy-Pontoise Cedex, France}
\\
\end{center}
\vskip 0.2cm

The nature of the deconfining phase transition in the 
2+1-dimensional SU(N) Georgi-Glashow model is investigated.
Within the dimensional-reduction hypothesis, 
the properties of the transition 
are described by a two-dimensional vectorial Coulomb gas models
of electric and magnetic charges.
The resulting critical properties are governed by a
generalized SU(N) sine-Gordon model with self-dual symmetry.
We show that this model displays a massless flow to 
an infrared fixed point which 
corresponds to the Z$_N$ parafermions conformal field theory.
This result, in turn, 
supports the conjecture of Kogan, Tekin, and Kovner that 
the deconfining transition in the
2+1-dimensional SU(N) Georgi-Glashow model belongs to 
the Z$_N$ universality class.


\section{Introduction}
\qquad
The 2+1 dimensional Georgi Glashow (GG) model
has attracted a lot of interest in the past
since it is a much simpler theory than QCD but still
retains some common interesting features like the existence of a
confinement phase.
The confinement phase of the GG model appears
in the weak-coupling limit and can be investigated
analytically. In particular, it has been
shown by Polyakov \cite{polyakov} that
the resulting phase
is a Coulomb plasma of monopoles and
antimonopoles
and the photon acquires a mass from the
Debye screening by monopoles.
The resulting phase is a
confinement phase since
a probe charge inserted in the vacuum will be screened
by monopoles \cite{kogan}.
The finite-temperature effect is an important issue
since confining gauge theories generally become
deconfined at high temperatures \cite{yaffe}.
The nature of the confinement-deconfinement phase
transition in the 2+1 dimensional GG model
has been analysed in detail \cite{agasyan,dunne,kovchegov,kogan}.
In particular, as first stressed by the authors
of Ref. \cite{dunne}, the massive W$^{\pm}$ gauge
bosons play  a crucial role for the deconfinement transition.
This phase transition stems from the competition 
between the monopoles and W$^{\pm}$ gauge
bosons which act as U(1) vortices.
Within the dimensional-reduction hypothesis, 
it has been 
shown that this competition results in an Ising critical
behavior \cite{dunne,kovchegov}.

The deconfinement transition in the 
SU(N)-generalization of the GG model
has also been investigated similarly \cite{kogansun}. 
As it will be briefly reviewed in Section 2,
within the dimensional-reduction hypothesis,
the low-energy Hamiltonian density, which governs
the resulting phase transition, takes
the form of a generalized
two-dimensional sine-Gordon model: 
\begin{eqnarray}
{\cal H}_N &=& \frac{1}{2}\left[\left(\partial_x {\vec \Phi} \right)^2
+ \left(\partial_x {\vec \Theta} \right)^2 \right]
\nonumber \\
&-& g \sum_{{\vec \alpha} \epsilon \Delta_+}\left[
\nord \cos\left( \sqrt{4\pi} {\vec \alpha} \cdot {\vec \Phi}\right)
\nord
+
\nord \cos\left( \sqrt{4\pi} {\vec \alpha} \cdot {\vec \Theta}\right)
\nord \right],
\label{sunsdsgham}
\end{eqnarray}
where the summation over ${\vec \alpha}$ is taken over the positive roots
of SU(N) normalized to unity (${\vec \alpha}^2 =1$),
and $\nord \nord$ denotes the normal ordering symbol.
The bosonic vector field ${\vec \Phi} = (\Phi_1,
\ldots, \Phi_{N-1})$ is made of
$N-1$ free boson fields
with chiral components $\Phi_{a R,L}$:
$\Phi_a = \Phi_{a L} + \Phi_{a R}$, ($a=1,..,N-1$).
The dual vector field ${\vec \Theta} = (\Theta_1,
\ldots, \Theta_{N-1})$ is defined by:
$\Theta_a = \Phi_{a L} - \Phi_{a R}$.
Model (\ref{sunsdsgham}) is a generalization
of the sine-Gordon for the boson vector field  ${\vec \Phi}$
with an additional pertubation
depending on the dual field  ${\vec \Theta}$.
This field theory has been introduced in Ref. \cite{boyanovsky}
for exploring critical properties of vectorial Coulomb
gas models of electric and magnetic charges.
The interacting part of model (\ref{sunsdsgham}) is a strongly
relevant perturbation with scaling dimension one and
its special structure makes it invariant under
the Gaussian duality symmetry:
${\vec \Phi} \leftrightarrow {\vec \Theta}$,  i.e. the exchange of
electric and magnetic charges in the Coulomb
gas context.
In what follows, such model will be referred to
as the SU(N) self-dual sine-Gordon (SDSG) model.
This self-duality symmetry opens a possibility for
the existence of a critical point in the infrared (IR) limit
which governs the deconfinement transition in the 
2+1 dimensional SU(N) GG model.
From the renormalization group (RG) point of view,
the SU(N) SDSG model (\ref{sunsdsgham})
will then be characterized by a massless flow from the ultraviolet (UV) fixed
point with central charge $c_{UV} = N -1$ to a conformally
invariant IR fixed point
with a smaller central charge $c_{IR} < N - 1$ according
to the c-theorem \cite{cth}.
The perturbative study of model (\ref{sunsdsgham}) has been
done in Refs. \cite{boyanovsky,sierra} and a fixed point has been found
whose nature is beyond the scope of these investigations.
In Ref. \cite{kogansun} in connection to the 
deconfinement transition in the 2+1 dimensional SU(N) GG model,
it has been conjected that the  IR fixed point
belongs to the Z$_N$ parafermions universality class
which is a conformally invariant 
theory (CFT) with central charge $c = 2(N -1)/(N+2)$ 
\cite{zamolo}.
In the following, we shall prove this conjecture 
in the general $N$ case and show that
the SU(N) SDSG model (\ref{sunsdsgham}) with $g>0$
displays a massless flow  to a Z$_N$ parafermionic fixed point.

The letter is organized as follows.
In the next section, we briefly review 
the connection between the SU(N) SDSG model (\ref{sunsdsgham})
and the phase transition in  the
SU(N) GG model. In Section 3, the emergence of the 
Z$_2$ criticality in Eq. (\ref{sunsdsgham}) for 
$N=2$ is reviewed for completeness and to fix
the notations.
We present then a proof of the conjecture for 
the first non-trivial case i.e. $N=3$ in Section 
4. Finally, we consider the general $N$ case
in the last section.

\section{Deconfinement transition in the 
SU(N) GG model} 
\qquad
The SU(N) GG model describes a SU(N) gauge theory
which interacts with a Higgs field
transforming in the adjoint representation.
Its Lagrangian density in the Euclidean reads 
as follows:
\begin{equation}
{\cal L}_N = \frac{1}{2} {\rm Tr} \left( F^2_{\mu \nu}
\right) + {\rm Tr} \left(\left( D_{\mu} \Phi \right)^2
\right) + V\left( \Phi \right) ,
\label{GGN}
\end{equation}
with 
$F_{\mu \nu} = \partial_{\mu} A_{\nu}
- \partial_{\nu} A_{\mu} + g \left[A_{\mu}, A_{\nu} \right]$,
$D_{\mu} \Phi = \partial_{\mu} \Phi
 + g \left[A_{\mu}, \Phi \right]$, 
$A_{\mu}  = A_{\mu}^A T^A$ and
$\Phi = \Phi^A T^A$ ($A=1,2, \ldots, N^2 - 1$),
$T^A$ being the generators of the Lie algebra of SU(N)
normalized as: $ {\rm Tr}(  T^A  T^B) = \delta^{A B}/2$.
In Eq. (\ref{GGN}),
the Higgs potential is supposed to be such that
the SU(N) gauge symmetry is spontaneously broken down to
U(1)$^{N-1}$.
In addition to the
Higgs field, the perturbative spectrum consists of $N-1$ 
massless photons 
and $N(N-1)$ massive gauge bosons in correspondence 
with the ladder operator $E_{\vec \alpha}$ of the Cartan-Weyl basis
of SU(N) 
(${\vec \alpha}$ being
the roots of SU(N) normalized to one) \cite{kogansun}.
In this basis, the Higgs vacuum-expectation value is diagonal:
$\langle \Phi \rangle = {\vec h} \cdot {\vec H}$, 
${\vec H}$ being the Cartan generators of SU(N).
The W bosons have the mass:
$m_{\rm W}
\equiv m_ {{\vec \alpha}} = g |{\vec h} \cdot {\vec \alpha}|$,
and carry the U(1)$^{N-1}$ (electric) charge:
${\vec e}_{\vec \alpha} = g \;{\vec \alpha}$.
In the weak-coupling regime, which is defined by $m_{\rm W} \sim
m_{\rm H} \gg g^2$, the massive-gauge bosons and the Higgs
field decouple at low-energy and a massless free gauge theory
remains.
However,
non-perturbative configurations (monopoles or instantons) give a
mass $m_{\gamma}$ to these photons.
Indeed, as is well known, model (\ref{GGN}) admits
stable classical solutions with finite action
($\Pi_2(SU(N)/U(1)^{N-1}) = Z_N$ for $N > 2$).
The magnetic field of these monopoles is:
$B^{\mu} = {\vec g} \cdot {\vec H} x^{\mu}/4\pi r^3$,
where the magnetic charge (${\vec g}$) satisfies the
condition \cite{englert}:
${\vec g} = 4 \pi \sum_{a=1}^{N-1}
n_a {\vec \beta}_a^{*}/g$,
${\vec \beta}_a^{*}$ being the dual simple roots
(${\vec \beta}_a^{*} ={\vec \beta}_a/|{\vec \beta}_a| =
{\vec \beta}_a$, ${\vec \beta}_a$ being the simple
roots of SU(N)) and $n_a$ are integers.

The weak-coupling phase corresponds to
a confined phase and a deconfinement transition
should occur at sufficiently high temperature.
The central question is 
the universality class of this 
transition in the general $N$ case.
In the regime $m_{\gamma} \ll T \ll m_{\rm W}$,
one can adopt the dimensional-reduction hypothesis
for exploring the phase transition since 
the size of the compactified direction ($T^{-1}$) is much
smaller than the average distance between the 
monopoles.
The monopoles form thus a two-dimensional
Coulomb gas with vectorial magnetic charges.
The contribution of the massive gauge bosons
is crucial for the physics
of the deconfinement as first stressed in Refs.
\cite{dunne,kogansun}.
In the following, 
we shall neglect the contribution
of the Higgs field for investigating
the deconfinement transition (see
Ref. \cite{antonovhiggs} for the
influence of a Higgs-boson mass).
The partition function, which describes
the two-dimensional vectorial Coulomb gas of
monopoles and massive gauge bosons, 
reads then as follows:
\begin{eqnarray}
{\cal Z} &=& \sum_{M,N = 0}^{\infty} \frac{1}{M ! N !}
\prod_{i=1}^{M}  \prod_{j=1}^{N}
\sum_{{\vec \alpha}_i}
\sum_{{\vec \alpha}_j}
\zeta_{{\vec \alpha}_i} {\tilde \zeta}_{{\vec \alpha}_j}
\nonumber \\
&& \int d^2 {\vec x}_i d^2 {\vec y}_j
\; \; \exp \left(
- {\cal S}\left({\vec x}_i, {\vec g}_{{\vec \alpha}_i};
{\vec y}_j, {\vec e}_{{\vec \alpha}_j}
\right) \right),
\label{partitionfunctionsun}
\end{eqnarray}
where ${\cal S}\left({\vec x}_i, {\vec g}_{{\vec \alpha}_i};
{\vec y}_j, {\vec e}_{{\vec \alpha}_j} \right)$
is the effective action of $M$ monopoles located at ${\vec x}_i$
with magnetic charges
${\vec g}_{{\vec \alpha}_i}$, fugacity $\zeta_{{\vec \alpha}_i}$
and $N$ W bosons at positions ${\vec y}_j$ with
electric charges ${\vec e}_{{\vec \alpha}_j}$,
fugacity ${\tilde \zeta}_{{\vec \alpha}_j}$.
This effective action has been carefully derived
in Ref. \cite{kogansun} but can also be found
in a phenomenological manner as in the SU(2) case \cite{kovchegov}.
It separates into three different parts:
the actions of two-dimensional Coulomb gas for the
monopoles and W bosons with an 
UV cut-off $T^{-1}$ and an interaction between
them which takes the form of an Aharonov-Bohm phase factor:
\begin{eqnarray}
{\cal S} &=& - \frac{T}{2\pi} \sum_{i < j}
{\vec g}_{{\vec \alpha}_i}
\cdot  {\vec g}_{{\vec \alpha}_j}
 \ln \left( T |{\vec x}_i - {\vec x}_j| \right)
\nonumber \\
&-&  \frac{1}{2\pi T} \sum_{i < j}
{\vec e}_{{\vec \alpha}_i}
\cdot  {\vec e}_{{\vec \alpha}_j}
 \ln \left( T |{\vec y}_i - {\vec y}_j| \right)
- \frac{i}{2\pi} \sum_{i=1}^{M} \sum_{j=1}^{N}  {\vec g}_{{\vec \alpha}_i}
\cdot  {\vec e}_{{\vec \alpha}_j}
\theta\left({\vec x}_i - {\vec y}_j \right),
\label{actiontotalsun}
\end{eqnarray}
with the neutral condition: $\sum_i {\vec e}_{{\vec \alpha}_i} =
\sum_j {\vec g}_{{\vec \alpha}_j} = {\vec 0}$, and 
$\theta\left({\vec x}_i - {\vec y}_j \right)$
is the angle between the vector connecting the monopole
at ${\vec x}_i$ and the W boson at ${\vec y}_j$
and a chosen spatial direction.
We then introduce a free massless bosonic vector
field ${\vec \Phi} = (\Phi_1, \ldots, \Phi_{N-1})$
and its dual field
${\vec \Theta} = (\Theta_1, \ldots, \Theta_{N-1})$
to express partition function (\ref{partitionfunctionsun})
in terms of these bosonic fields:
\begin{eqnarray}
{\cal Z} &\sim& \sum_{M,N = 0}^{\infty} \frac{1}{M ! N !}
\prod_{i=1}^{M}  \prod_{j=1}^{N}
\sum_{{\vec \alpha}_i}
\sum_{{\vec \alpha}_j}
\zeta_{{\vec \alpha}_i} {\tilde \zeta}_{{\vec \alpha}_j}
\int d^2 {\vec x}_i d^2 {\vec y}_j
\nonumber \\
&& \bigg  \langle \prod_{i=1}^{M} \exp \left( i
\sqrt{T} {\vec g}_{{\vec \alpha}_i}
\cdot
{\vec \Phi} \left({\vec x}_i\right) \right)
\prod_{j=1}^{N} \exp \left( i T^{-1/2} {\vec e}_{{\vec \alpha}_j}
\cdot
{\vec \Theta} \left({\vec y}_j\right) \right)
\bigg \rangle .
\label{partsunbosonized}
\end{eqnarray}
The effective
Hamiltonian density, which describes
the deconfinement-confinement transition
of the SU(N) GG model (\ref{GGN}),
can then be deduced by performing 
the summations in Eq. (\ref{partsunbosonized}):
\begin{eqnarray}
{\cal H}_{\rm eff} &=&
\frac{1}{2}\left[\left(\partial_x {\vec \Phi} \right)^2
+ \left(\partial_x {\vec \Theta} \right)^2 \right]
\nonumber \\
&-& \sum_{{\vec \alpha} \epsilon \Delta_+}\left[
\zeta_{{\vec \alpha}}
 \cos\left( \frac{4 \pi\sqrt{T}}{g}
\; {\vec \alpha} \cdot {\vec \Phi}\right)
+
{\tilde \zeta}_{{\vec \alpha}}
 \cos\left(\frac{g}{\sqrt{T}}
\; {\vec \alpha} \cdot {\vec \Theta}\right)
 \right],
\label{sunGGhameff}
\end{eqnarray}
where the summation is taken over the positive
roots of SU(N) and the fugacities have been rescaled. 
This low-energy effective theory has been first derived
in Ref. \cite{kogansun} by means of a similar Coulomb-gas
analysis and also by using
the magnetic Z$_N$ symmetry
which is spontaneously broken in the confinement phase \cite{kovner}.
Model (\ref{sunGGhameff}) is a generalization of the sine-Gordon model
for multi-boson fields  and
describes the competition between monopoles and vortices
in this SU(N) GG model.
The one-loop RG equations for this model have been investigated in
Refs. \cite{boyanovsky,sierra} and they are quite complex in general.
However, as in Ref. \cite{kogansun}, we shall consider
here a simpler case, which is stable under the RG flow,
where all monopole fugacities
are equal $\zeta_{\vec \alpha} =
\zeta$ and similarly for the
vortex fugacities: ${\tilde \zeta}_{\vec \alpha} = {\tilde \zeta}$.
A non-trivial stable IR fixed point has been found
perturbatively within this manifold
\cite{boyanovsky,sierra} which should govern
the deconfinement transition
in the SU(N) GG model (\ref{GGN}) \cite{kogansun}.
The low-energy physics of the resulting model
can be deduced qualitatively by simple scaling arguments.
The scaling dimensions of the
two vertex operators in Eq. (\ref{sunGGhameff})
are: $\Delta = T 4 \pi/ g^2$
and ${\tilde \Delta} =  g^2/T 4 \pi$, i.e. $\Delta {\tilde \Delta} = 1$.
When $T < g^2/8 \pi$, 
one has $\Delta < 1/2$ and ${\tilde \Delta} > 2$:
the perturbation depending on the dual vector field is irrelevant
whereas the monopole term is a strongly relevant perturbation.
The low-energy theory reduces to a sine-Gordon model
for the ${\vec \Phi}$ field
and a mass-gap is induced.
It corresponds to the confinement phase where the massive
W gauge bosons can be neglected.
At high-temperature $T > g^2/2 \pi$,  the monopole
term is now irrelevant $\Delta > 2$ and
model (\ref{sunGGhameff})
reduces again to a sine-Gordon model but for the
dual vector field ${\vec \Theta}$
with a relevant perturbation (${\tilde \Delta} < 1/2$).
A mass-gap is still present 
but it corresponds to the deconfined phase
with the unbinding of the W gauge bosons.
The transition between these two phases
takes place in the regime: $g^2/8 \pi < T < g^2/2 \pi$ where
both pertubations of Eq. (\ref{sunGGhameff}) are relevant.
On general grounds, the transition is expected
to appear along the self-dual
line where model (\ref{sunGGhameff}) is invariant under
the duality transformation ${\vec \Phi} \leftrightarrow {\vec \Theta}$,
i.e. the symmetry between electric and magnetic charges.
This self-dual symmetry is realized when $\zeta = {\tilde \zeta}$
and $T = g^2/4\pi$.
In this case, the confinement-deconfinement transition is thus
governed by the SU(N) SDSG model (\ref{sunsdsgham}).
In Ref. \cite{kogansun}, it has been conjectured
that this phase transition belongs to the Z$_N$
universality class corresponding to the Z$_N$
parafermionic CFT \cite{zamolo}.
We now turn to the proof of this conjecture. 

\section{Ising criticality}

\qquad
We start with the simplest
case, i.e. $N=2$, where
the SU(2) SDSG model takes the form:
\begin{equation}
{\cal H}_{2} = \frac{1}{2} \left[
\left(\partial_x \Phi\right)^2 +
\left(\partial_x \Theta\right)^2 \right] -
g\left[ \nord \cos\left(\sqrt{4\pi} \; \Phi\right) \nord
+ \nord \cos\left(\sqrt{4\pi} \; \Theta\right) \nord \right].
\label{ham4pi}
\end{equation}
This model is well known (see e.g. Ref. \cite{ogilvie})
and can be exactly diagonalized even in a more general case 
when the two cosine
terms in Eq. (\ref{ham4pi}) have independent amplitudes.
Assuming that the boson field $\Phi$
is compactified
with radius $R = 1/\sqrt{4\pi}$ or
$R = 1/\sqrt{\pi}$, i.e. the Dirac point (see the book \cite{dms}
for a review)
in our notation, the model can be
refermionized by introducing two Majorana fermions fields $\xi^{1,2}$.
This procedure is nothing but the standard bosonization of two Ising
models \cite{zuber,boyanovskyising,bookboso}.
The bosonization rules are given by
\begin{eqnarray}
\xi_R^{1} + i \xi_R^{2} &=& \frac{1}{\sqrt{\pi}} :\exp\left(i\sqrt{4\pi}
\Phi_R\right):, \nonumber \\
\xi_L^{1} + i \xi_L^{2} &=& \frac{1}{\sqrt{\pi}} :\exp\left(-i\sqrt{4\pi}
\Phi_L\right):,
\label{bosoising}
\end{eqnarray}
where $\Phi_{R,L}$ are the chiral components of the Bose field:
$\Phi_L = (\Phi + \Theta)/2$ and $\Phi_R = (\Phi - \Theta)/2$.
They satisfy $[\Phi_R, \Phi_L ] = i/4$ to insure
the anticommutation relation between right and left fermions.
One can easily check that the bosonic representation
(\ref{bosoising}) is consistent with the defining operator
product expansion (OPE) for the Majorana fields:
\begin{equation}
\xi_L^a\left(z\right) \xi_L^b\left(w\right)
\sim \frac{\delta^{ab}}{2\pi\left(z-w\right)},
\label{majope}
\end{equation}
with a similar OPE for the right Majorana fermion.
The self-dual Hamiltonian (\ref{ham4pi}) can then be expressed in terms
of these Majorana fermions:
\begin{equation}
{\cal H}_{2} = -\frac{i}{2} \sum_{a=1}^{2}\left(
\xi_R^{a} \partial_x \xi_R^{a} -
\xi_L^{a} \partial_x \xi_L^{a} \right)
- i m \xi_R^{2} \xi_L^{2},
\label{ham4pifer}
\end{equation}
with $m = 2  \pi g$. The Hamiltonian
of the SU(2) SDSG model separates thus into two commuting pieces.
One of the decoupled degrees of freedom corresponds to an effective
off-critical Ising model described by the massive Majorana
fermion $\xi_{R,L}^2$, whereas
the second Majorana field $\xi_{R,L}^1$ remains massless.
The Gaussian self-dual symmetry $\Phi \leftrightarrow \Theta$
coincides to the Kramers-Wannier (KW) duality symmetry of
the Ising model associated to the Majorana
fermion $\xi^1$: no mass term $i \xi_R^{1} \xi_L^{1}$, which
is odd under the KW duality, can appear in the effective
Hamiltonian.
The existence of this massless Majorana mode signals the Z$_2$
(Ising) criticality of the SU(2) SDSG model (\ref{ham4pi})
and the Ising nature of the deconfining transition 
in the SU(2) GG model \cite{dunne,kovchegov}.

\section{Z$_3$ criticality}

\qquad
We now turn to the $N=3$ case which is much more complex
than the previous case. In this section, we shall 
argue that the SU(3) SDSG model 
displays in the IR limit a Z$_3$ critical behavior
with central charge $c=4/5$.
To this end, let us first rewrite model (\ref{sunsdsgham})  explicitely
in terms of the two bosonic fields using
the roots of SU(3):
\begin{eqnarray}
{\cal H}_3 &=& \frac{1}{2}\left[\left(\partial_x {\vec \Phi} \right)^2
+ \left(\partial_x {\vec \Theta} \right)^2 \right]
- g \left[
\nord \cos\left( \sqrt{4\pi} \Phi_s\right)
+
\cos\left( \sqrt{\pi} \Phi_s\right)
\cos\left(\sqrt{3 \pi} \Phi_f\right) \nord
\right.
\nonumber \\
&+& \left. \nord \cos\left( \sqrt{4\pi} \Theta_s\right) +
\cos\left( \sqrt{\pi} \Theta_s\right)
\cos\left(\sqrt{3 \pi} \Theta_f\right)
\nord \right],
\label{su3sdsgham}
\end{eqnarray}
where ${\vec \Phi} = (\Phi_s, \Phi_f)$
and ${\vec \Theta} = (\Theta_s, \Theta_f)$ for $N=3$.
In the following, we shall assume that $\Phi_s$
(respectively $\Phi_f$) is a bosonic field
compactified with radius
$R_s = 1/ \sqrt{\pi}$ (respectively
$R_f =  \sqrt{3 /\pi}$).
As in Section 3, at this free-fermion point
for the bosonic field $\Phi_s$,
one can refermionize model (\ref{su3sdsgham}) by introducing
two Majorana fermions $\xi^{1,2}$ using Eq. (\ref{bosoising}).
We need also the refermionization of the vertex operators
$\cos(\sqrt{\pi} \Phi_s)$ and $\cos(\sqrt{\pi} \Theta_s)$
with scaling dimension $1/4$.
They can be expressed in terms of the
Ising order ($\sigma_{1,2}$) and disorder
($\mu_{1,2}$) parameters
of the two underlying Ising models 
\cite{boyanovskyising,bookboso}:
$\nord \cos(\sqrt{\pi} \Phi_s) \nord
\sim \mu_1 \mu_2$ and
$\nord \cos(\sqrt{\pi} \Theta_s) \nord
\sim \sigma_1 \mu_2$.
Model (\ref{su3sdsgham}) can then be recasted in terms
of this Ising$^2$ CFT and the free massless bosonic
field $\Phi_f$:
\begin{eqnarray}
{\cal H}_3 &=&
\frac{1}{2}\left[\left(\partial_x \Phi_f \right)^2
+ \left(\partial_x \Theta_f \right)^2 \right]
- \frac{i}{2}\sum_{a=1}^{2}\left(\xi_R^a \partial_x
\xi_R^a -
\xi_L^a \partial_x
\xi_L^a \right)
\nonumber \\
&-& i m \xi_R^2 \xi_L^2
- g \left[
\mu_1 \mu_2
\nord \cos\left(\sqrt{3 \pi} \Phi_f\right) \nord
+
\sigma_1 \mu_2
\nord
\cos\left(\sqrt{3 \pi} \Theta_f\right)
\nord \right],
\label{su3sdsghamref}
\end{eqnarray}
with $m = 2 \pi g$. Clearly due to the mass
term in Eq. (\ref{su3sdsghamref}),
the Ising model, corresponding to the fermion
$\xi^2$, is out of criticality and since $m >0$ ($g >0$)
it belongs to its high-temperature phase in our convention
($m \sim T - T_c$).
In this case, the Ising disorder operator condenses:
$\langle \mu_2 \rangle \ne 0$.
One can formally integrate out over this massive degrees
of freedom and rewrite model (\ref{su3sdsghamref})
in this low-energy limit ($E \ll m$):
\begin{eqnarray}
{\cal H}_3 &=&
\frac{1}{2}\left[\left(\partial_x \Phi_f \right)^2
+ \left(\partial_x \Theta_f \right)^2 \right]
- \frac{i}{2}\left(\xi_R^1 \partial_x
\xi_R^1 -
\xi_L^1 \partial_x
\xi_L^1 \right)
\nonumber \\
&- \lambda& \left[
\mu_1
\nord \cos\left(\sqrt{3 \pi} \Phi_f\right) \nord
+
\sigma_1
\nord
\cos\left(\sqrt{3 \pi} \Theta_f\right)
\nord \right],
\label{su3eff}
\end{eqnarray}
where $\lambda$ is a non-universal constant that results
from the integration of the massive degrees of
freedom. The effective Hamiltonian  (\ref{su3eff}) describes
a critical Ising model and a free massless boson field
that interact with a strongly relevant perturbation
with scaling dimension $7/8$.

The next step of the approach is to switch on
a different basis to determine the main
effect of the relevant perturbation of
Eq. (\ref{su3eff}).
To this end, it is important to observe that the free massless bosonic
field $\Phi_f$ has a very special radius
$R_f =  \sqrt{3 /\pi}$ in the classification
of the CFT with central charge $c=1$.
At this radius, it displays a CFT with an extended
symmetry: a $N=2$ (respectively $N=1$) superconformal field theory
(SCFT)
whether the bosonic field $\Phi_f$ is compactified
along a circle (respectively an orbifold) \cite{waterson}.
The conformal symmetry of the UV fixed point of
model (\ref{su3eff}), i.e. Ising $\times$ [$c=1$ SCFT],
with central charge $c=3/2$
can also be described in terms of the product
of TIM $\times$ Potts CFTs, where the TIM and Potts
refers respectively to the tricritical Ising and three-state
Potts CFTs.
The precise conformal embedding has been derived
by the authors of Ref. \cite{bul} and the result
depends on the nature of the compactification
of the bosonic field $\Phi_f$:
\begin{eqnarray}
{\rm Ising} \times \left(c=1 \; \; N=1 \; \; {\rm SCFT} \right)
&=& P\left[{\cal M}_4 \times {\cal M}_5 \right]
\label{embeddingscft1}
\\
{\rm Ising} \times \left(c=1 \; \; N=2 \; \; {\rm SCFT} \right)
&=& P\left[\left(c= 7/10 \; \; N=1 \; \; {\rm SCFT} \right)
 \times {\rm Z}_3 \right]  ,
\label{embeddingscft2}
\end{eqnarray}
where ${\cal M}_p$ denotes the minimal model
series with central charge $c_p = 1 - 6/p(p+1)$ i.e. the
TIM and Potts CFTs for $p=4$ and $p=5$ respectively
\cite{dms}.
The $c= 7/10 \; \; N=1 \; \; {\rm SCFT}$ is known
to be equivalent to the TIM CFT \cite{friedan}.
In Eqs. (\ref{embeddingscft1},\ref{embeddingscft2}), a
projection $P$ is crucial to realize the equivalence as demonstrated
in Ref. \cite{bul}.
For instance,
if we denote the primary fields of the ${\cal M}_p$ CFT
as $\Phi^{(p)}_{r,s}$ ($ 1 \le r \le p-1, 1 \le s \le p$)
then the projection $P$, for the equivalence
of Eq. (\ref{embeddingscft1}), restricts the operators
to the subset: $\{\Phi^{(4)}_{r,s} \Phi^{(5)}_{s,q} \}$
\cite{bul}.
By looking at the dimensions of the fields in this subset and
OPEs consistency, we find that the original perturbation with
scaling dimension $7/8$ of Eq. (\ref{su3eff})  identifies to
the submagnetic operator $\sigma^{'}$
of the TIM CFT
with scaling dimension $7/8$:
\begin{equation}
\sigma^{'} \sim
\;
\mu_1
\nord \cos\left(\sqrt{3 \pi} \Phi_f\right) \nord
+
\sigma_1
\nord
\cos\left(\sqrt{3 \pi} \Theta_f\right)
\nord .
\label{submagbos}
\end{equation}

With all these results, we can now express
the low-energy effective Hamiltonian
(\ref{su3eff}) in the new basis:
\begin{equation}
{\cal H}_{3} = {\cal H}^{{\rm Z}_3}_{0} + {\cal H}^{{\rm TIM}}_{0}
- \lambda \sigma^{'} ,
\label{h3effnew}
\end{equation}
where ${\cal H}^{{\rm Z}_3}_{0}$ (respectively ${\cal H}^{{\rm TIM}}_{0}$)
denotes the Hamiltonian of the three-state Potts (respectively TIM) CFT.
The deformation of the TIM CFT by the subleading magnetization $\sigma^{'}$
(i.e. $\Phi^{\rm TIM}_{2,1}$) is known to be an integrable massive field theory
\cite{mussardo}. Therefore, we deduce that
the SU(3) SDSG model 
flows in the far IR limit towards a fixed point with central charge $c=4/5$
corresponding to the Z$_3$ universality class.
In this respect, we observe that the
original Gaussian self-duality ${\vec \Phi}
\leftrightarrow {\vec \Theta}$ of model (\ref{su3sdsgham})
coincides with the KW symmetry
of the three-state Potts model.
In addition, it might be interesting to see
the emergence of this Potts criticality
from another point of view.
Indeed, it is possible to rewrite model (\ref{su3sdsghamref})
before the integration
of the massive Majorana fermion $\xi^{2}$ in the Z$_3$ $\times$
TIM basis
using the identification (\ref{submagbos}):
\begin{equation}
{\cal H}_{3} =
- \frac{i}{2}\left(\xi_R^2 \partial_x
\xi_R^2 -
\xi_L^2 \partial_x
\xi_L^2 \right) + {\cal H}^{{\rm Z}_3}_{0}
+ {\cal H}^{{\rm TIM}}_{0}
- i m \xi_R^2 \xi_L^2 - g \mu_2\sigma^{'} .
\label{h3ini}
\end{equation}
The Ising and TIM CFTs can be combined to form another CFT
with extended symmetry \cite{bul,affleck}: a ${\cal W}_3$ CFT with
Z$_3$ symmetry which has been introduced by
Fateev and Zamolodchikov \cite{w3}.
More precisely, the $P({\rm Ising} \times {\rm TIM})$ CFT corresponds
to the Z$_3^{[5]}$ CFT with central charge $c=6/5$ \cite{bul}.
Some character decompositions have been
found in Ref. \cite{affleck}
and we have:
\begin{equation}
\chi_{1/16}^{{\rm Ising}} \chi_{7/16}^{{\rm TIM}} =
\chi_{0}^{{\rm Ising}}
\chi_{3/2}^{{\rm TIM}}
+ \chi_{1/2}^{{\rm Ising}}
\chi_{0}^{{\rm TIM}} = \chi_{1/2}^{{\cal W}_3},
\label{characteridentw3}
\end{equation}
where $\chi_h^{\rm CFT}$ is the character of the conformal tower with
holomorphic weight $h$ of the underlying CFT.
We have checked numerically
the identities (\ref{characteridentw3}) with Mathematica.
The primary fields of  the Z$_3^{[5]}$ CFT
can be noted as $\Phi^{{\rm W}_3}_{(\lambda, \lambda^{'})}$,
where $\lambda$ and $\lambda^{'}$ are dominant weights of SU(3) algebra i.e.
$\lambda = \sum_{i=1}^{2} (l_i - 1) \lambda_i$,
$\lambda^{'} = \sum_{i=1}^{2} (l^{'}_i - 1) \lambda_i$
($\lambda_i$ being the
fundamental weights of SU(3) and $l_i, l^{'}_i$ are positive).
The Z$_3^{[5]}$ primary field with scaling dimension
one is $\Phi^{{\rm W}_3}_{(\lambda_1+ \lambda_2,0)}$ which
transform in the adjoint and trivial representations of SU(3).
The relation (\ref{characteridentw3}) leads us to
expect that model (\ref{h3ini})
can be written in the following compact form:
\begin{equation}
{\cal H}_{3}  =
{\cal H}^{{\rm Z}_3}_{0} + {\cal H}^{{\rm W}_3}_{0}
- \lambda \Phi^{{\rm W}_3}_{(\lambda_1+ \lambda_2,0)} ,
\label{h3ininew}
\end{equation}
where ${\cal H}^{{\rm W}_3}_{0}$  is the Hamiltonian of the Z$_3^{[5]}$ CFT.
It is interesting to note that some
integrable deformations of the ${\cal W}_3$ CFT
have been found in Ref. \cite{devega}.
In particular, a Z$_3^{[5]}$ CFT perturbed by a primary
field which transform
in the adjoint and trivial representations
of SU(3) is a massive integrable field theory.
Therefore, we conclude again
that the SU(3) SDSG model 
flows towards the Z$_3$ fixed point in the IR limit
within this approach.  In this respect, we deduce that 
the deconfining transition in the SU(3) GG model belongs
to the Z$_3$ universality class in full agreement
with the conjecture of Kogan {\it et al.} \cite{kogansun}.

\section{The general case}

\qquad
We now consider the general $N$
case namely that the SU(N) SDSG 
flow towards the Z$_N$ fixed point in the IR limit.
In this respect,
let us introduce $N$ copies of the SU(2)$_1$
Wess-Zumino-Novikov-Witten (WZNW) CFT and
consider the following Hamiltonian density:
\begin{equation}
{\cal H} = \frac{2 \pi}{3} \sum_{a=1}^{N}
\left( \nord {\vec J}_{a R}^2 \nord
+ \nord {\vec J}_{a L}^2 \nord \right)
- \frac{g}{2} \sum_{a < b}\left(
{\rm Tr} g_a {\rm Tr} g_b
- {\rm Tr}\left(g_a {\vec \sigma}\right)
\cdot {\rm Tr} \left(g_b {\vec \sigma}\right)
\right) ,
\label{WZWmodels}
\end{equation}
where ${\vec J}_{a R,L}$ are the chiral SU(2)$_1$
current of the ath WZNW CFT ($a=1,\ldots,N$) and
${\vec \sigma}$ are the Pauli matrices.
These currents satisfy the SU(2)$_1$ current algebra
defined by \cite{bookboso}:
\begin{eqnarray}
J^{\alpha}_{a L} \left(z\right)
J^{\beta}_{b L} \left(\omega \right) \sim
\frac{\delta^{\alpha \beta} \delta_{ab}}{8 \pi^2 \left(z - \omega\right)^2}
+
\frac{i\delta_{ab}\epsilon^{\alpha \beta \gamma}
J^{\gamma}_{a L} \left(\omega \right)}{2 \pi
\left(z - \omega\right)},
\label{curope}
\end{eqnarray}
with $\alpha, \beta , \gamma = x,y,z$ and 
a similar OPE for the right
current. 
The interacting part of model (\ref{WZWmodels}),
that we denote ${\cal H}_{\rm int}$,
is a strongly relevant perturbation with scaling
dimension one which is made of the WZNW
field $g_a$. This primary field transforms in the
fundamental representation of SU(2) and is defined
by the following OPEs \cite{dms}:
\begin{eqnarray}
J^{\alpha}_{a L} \left(z\right) g_b \left(\omega ,
\bar \omega \right) &\sim&
- \frac{\delta_{ab}}{2 \pi
\left(z - \omega\right)} \; \;
\frac{\sigma^{\alpha}}{2} \; g_b \left(\omega ,
\bar \omega \right)
\nonumber \\
J^{\alpha}_{a R} \left(\bar z\right) g_b \left(\omega ,
\bar \omega \right) &\sim&
\frac{\delta_{ab}}{2 \pi
\left(\bar z - \bar \omega\right)}
\; \; g_b \left(\omega ,
\bar \omega \right) \;
\frac{\sigma^{\alpha}}{2} .
\label{wznwgope}
\end{eqnarray}
It might be interesting to note that model (\ref{WZWmodels}) appears in the context 
of coupled electronic chains \cite{tsvelik}.
The next step of the approach
is to consider the following conformal embedding to analyse
the main effect of ${\cal H}_{\rm int}$:
\begin{equation}
SU(2)_1 \times SU(2)_1 \times ..
\times SU(2)_1 \rightarrow SU(2)_N \times {\cal G}_N,
\label{embedgenn}
\end{equation}
where SU(2)$_N$ is the level-N  SU(2) WZNW CFT with
central charge $c_N = 3N/(N+2)$ 
and ${\cal G}_N$ is a discrete CFT with
central charge $c_{{\cal G}_N}=N(N - 1)/(N+2)$.
The latter central charge coincides with the sum
of the central charge of the $N-1$ first minimal
models:
\begin{equation}
c_{{\cal G}_N} = \frac{N \left(N - 1 \right)}{\left(N+2\right)} =
\sum_{m=2}^{N+1} \left( 1 - \frac{6}{m\left(m+1\right)}
\right),
\label{idencharge}
\end{equation}
which leads us to expect that ${\cal G}_N$ should
be related to the product ${\cal M}_3 \times {\cal M}_4 \times \ldots
\times {\cal M}_{N+1}$.
The precise identification requires a projection P:
${\cal G}_N \sim P({\cal M}_3 \times {\cal M}_4 \ldots
\times {\cal M}_{N+1})$, which has been found in Ref.
\cite{bul}.
In fact, this projection restricts the product of the primary
fields of ${\cal M}_3 \times {\cal M}_4 \times \ldots
\times {\cal M}_{N+1}$ to the subset:
\begin{equation}
\{ \Phi^{(3)}_{r_3, s_3} \Phi^{(4)}_{s_3, s_4}
\Phi^{(5)}_{s_4, s_5} \ldots
\Phi^{(N)}_{s_{N-1}, s_N} \Phi^{(N+1)}_{s_N, s_{N+1}} \},
\label{projprimary}
\end{equation}
with $1 \le r_p \le p-1$ and
$1 \le s_p \le p$.
This quantum equivalence can be proved by a recursive
approach \cite{phle}.
We would like to express
${\cal H}_{\rm int}$ in the new basis
i.e. $SU(2)_N \times {\cal G}_N$.
To this end, we introduce the SU(2)$_N$
chiral currents ${\vec I}_{R,L}$ which are the sum
of the SU(2)$_1$ currents:
${\vec I}_{R,L} = \sum_{a=1}^{N} {\vec J}_{a R,L}$.
The key point of the analysis is that
${\cal H}_{\rm int}$ does not depend
on the SU(2)$_N$ CFT but expresses only
in terms of the fields of ${\cal G}_N$.
To show this, it is sufficient to determine
the OPE between the SU(2)$_N$ currents and ${\cal H}_{\rm int}$
and it should be zero.
This calculation can be done by using the defining OPEs
(\ref{wznwgope}) and we find:
\begin{eqnarray}
I^{\alpha}_{L} \left(z\right)
\left(
{\rm Tr} g_a {\rm Tr} g_b
- {\rm Tr}\left(g_a {\vec \sigma}\right)
\cdot {\rm Tr} \left(g_b {\vec \sigma}\right)
\right)  \left(\omega, \bar \omega \right) &\sim& 0
\nonumber \\
I^{\alpha}_{R} \left(\bar z\right)
\left(
{\rm Tr} g_a {\rm Tr} g_b
- {\rm Tr}\left(g_a {\vec \sigma}\right)
\cdot {\rm Tr} \left(g_b {\vec \sigma}\right)
\right)  \left(\omega, \bar \omega \right) &\sim& 0 ,
\label{opecal}
\end{eqnarray}
so that ${\cal H}_{\rm int}$ depends only
on the fields of ${\cal G}_N$.
We found an operator 
$\prod_{m=3}^{N} \Phi_{m-2,m-1}^{(m)} \Phi_{N-1,N+1}^{(N+1)}$ 
in the
${\cal G}_N \sim
P({\cal M}_3 \times {\cal M}_4 \ldots
\times {\cal M}_{N+1})$ CFT which couples
all minimal models involved with scaling dimension one:
\begin{equation}
\Delta = \sum_{m=3}^N \frac{3}{2 m \left(m +1 \right)}
+ \frac{N+4}{2 \left(N +1 \right)} = 1 .
\label{identdim}
\end{equation}
Therefore, due to the special
structure of ${\cal H}_{\rm int}$, a mass gap in the ${\cal G}_N$ sector
will be opened and
model (\ref{WZWmodels}) displays a massless flow
from the UV fixed with central charge $c_{\rm UV} = N$
to an IR fixed point with SU(2)$_N$ criticality:
$c_{\rm IR} = 3N/(N+2)$.

The connection with the SU(N) SDSG model
can then be made by exploiting the fact that
the SU(2)$_1$ CFT has a free-field representation
in terms of a massless bosonic field compactified
on a circle at the self-dual radius: $R = 1/\sqrt{2\pi}$
\cite{bookboso,dms}.
For the $N$ copies of the SU(2)$_1$ theory of the
original model (\ref{WZWmodels}), we thus introduce
the bosonic field ${\varphi}_a$ ($a=1,2, \ldots, N$) with its chiral
components ${\varphi}_{a R,L}$.
A faithfull representation of the OPEs
(\ref{wznwgope}) for  the WZNW tensor field $g_a$
is given by \cite{bookboso}: 
\begin{eqnarray}
g_a  = \frac{1}{\sqrt{2}}
\left(
\begin{array}{lccr}
\nord e^{-i \sqrt{2\pi} {\varphi}_a} \nord &
i \nord e^{-i \sqrt{2\pi} {\vartheta}_a} \nord \\
i \nord e^{i \sqrt{2\pi} {\vartheta}_a} \nord  &
\nord e^{i \sqrt{2\pi} {\varphi}_a} \nord
\end{array} \right) ,
\label{gbos}
\end{eqnarray}
from which we deduce the bosonic 
representation of model (\ref{WZWmodels}):
\begin{eqnarray}
{\cal H} &=&
\frac{1}{2} \sum_{a=1}^{N}
\left[ \left(\partial_{x}\varphi_{a R}\right)^2
+ \left(\partial_{x}\varphi_{a L}\right)^2
\right]
\nonumber \\
&-& g \sum_{a<b}
\left(
\nord \cos (\sqrt{2\pi}(\varphi_{a}-\varphi_{b})) \nord
+ \nord \cos (\sqrt{2\pi}(\vartheta_{a}-\vartheta_{b})) \nord
\right) .
\label{hambose}
\end{eqnarray}
We then perform a canonical transformation on
the Bose fields to simplify Eq. (\ref{hambose}).
To this end, we introduce a bosonic field
$\Phi_{cR,L}$ and $N-1$ other bosonic fields $\Phi_{l R,L}$
($l = 1, \ldots, N-1$) as follows:
\begin{eqnarray}
\Phi_{c R(L)} &=& \frac{1}{\sqrt{N}}\left(
\varphi_1 + \ldots + \varphi_N \right)_{R(L)}
\nonumber \\
\Phi_{l R(L)} &=& \frac{1}{\sqrt{l(l+1)}}\left(
\varphi_1 + \ldots + \varphi_l - l  \varphi_{l+1}\right)_{R(L)} .
\label{SUNbasis}
\end{eqnarray}
In this new basis, Hamiltonian (\ref{hambose}) reads:
\begin{eqnarray}
{\cal H} &=&
\frac{1}{2}\left[\left(\partial_x \Phi_c \right)^2
+ \left(\partial_x \Theta_c \right)^2 \right]
+
\frac{1}{2}\left[\left(\partial_x {\vec \Phi} \right)^2
+ \left(\partial_x {\vec \Theta} \right)^2 \right]
\nonumber \\
&-& g \sum_{{\vec \alpha} \epsilon \Delta_+}\left[
\nord \cos\left( \sqrt{4\pi} {\vec \alpha} \cdot {\vec \Phi}\right)
\nord
+
\nord \cos\left( \sqrt{4\pi} {\vec \alpha} \cdot {\vec \Theta}\right)
\nord \right],
\label{hambosefin}
\end{eqnarray}
where the summation over ${\vec \alpha}$ is taken over the positive roots
of SU(N),  ${\vec \Phi} = (\Phi_1, 
\ldots, \Phi_{N-1})$, and
${\vec \Theta} = (\Theta_1,
\ldots, \Theta_{N-1})$.
We thus observe that the initial model (\ref{WZWmodels})
separates into two commuting pieces:
${\cal H}  = {\cal H}_c^{0} + {\cal H}_N$,
a free-massless boson Hamiltonian ${\cal H}_c^{0}$
for the field $\Phi_c$ and the SU(N) SDSG model.
Since we know that model (\ref{WZWmodels}) admits
a massless flow onto the SU(2)$_N$ fixed point, we deduce
that the SU(N) SDSG model flows in the
IR limit to a conformally invariant fixed point with
central charge $c_{\rm IR} = 2(N-1)/(N+2)$.
The massless free-bosonic field $\Phi_c$, which decouples
from the interaction of Eq. (\ref{hambosefin}),
is compactified with radius $R_c = \sqrt{N/2\pi}$ and describes
an extended U(1)$_N$ rational $c=1$ CFT \cite{verlinde}.
Moreover, the Z$_N$ parafermionic CFT is known to be
equivalent to the coset: Z$_N \sim$ SU(2)$_N$/U(1)$_N$
\cite{zamolo,dms} so that we conclude that
the SU(N) SDSG model 
admits a massless flow onto the Z$_N$ CFT.
The original Gaussian self-duality ${\vec \Phi}
\leftrightarrow {\vec \Theta}$ coincides with
the KW self-duality symmetry of the Z$_N$
parafermionic model.
Finally, the massless flow found here implies that
the confinement-deconfinement phase transition
of the 2+1 dimensional SU(N) GG model (\ref{GGN})
should belong to the Z$_N$ universality class.
In this respect, this result demonstrates
the conjecture presented in Ref. \cite{kogansun}.
Other implications of the quantum equivalence approach (\ref{embedgenn})
will be discussed elsewhere \cite{phle}.

\section*{Acknowledgements}

\qquad
We would like to thank H. Saleur, E. Boulat, K. Totsuka, P. Azaria,
A. A. Nersesyan, E. Orignac, and A. M. Tsvelik for helpful discussions. 
The author is also grateful to
the Abdus Salam ICTP for hospitality when this work was 
initiated.

\end{document}